\documentclass[twocolumn,showpacs,preprintnumbers,amsmath,amssymb,pre,superscriptaddress]{revtex4}
\usepackage{color}

\usepackage{graphicx}
\usepackage{dcolumn}
\usepackage{bm}
\usepackage{subfigure}
\usepackage{amssymb}
\renewcommand{\vec}{\boldsymbol}

\begin{document}

\title{Effective coupling parameter for 2D Yukawa liquids \\and non-invasive measurement of plasma parameters }

\author{T. Ott}
\affiliation{%
    Christian-Albrechts-Universit\"at zu Kiel, Institut f\"ur Theoretische Physik und Astrophysik, Leibnizstra\ss{}e 15, 24098 Kiel, Germany
}%

\author{M. Stanley}%
\affiliation{%
Department of Physics, University of Cambridge, J. J. Thomson Avenue, Cambridge, United Kingdom
}%

\author{M. Bonitz}%
\affiliation{%
    Christian-Albrechts-Universit\"at zu Kiel, Institut f\"ur Theoretische Physik und Astrophysik, Leibnizstra\ss{}e 15, 24098 Kiel, Germany
}%

\date{\today}

\begin{abstract}
We derive an effective coupling parameter for two-dimensional Yukawa systems 
based on the height of the first maximum of the pair distribution function. Two variants --
one valid in the high-coupling range, the other for arbitrary couplings of the liquid --
are derived. Comparison to previous approaches to Yukawa coupling parameters shows that the present 
concept is more general and more accurate.
Using, in addition, dynamical information contained in the velocity autocorrelation 
function, we outline a reference data method that can be employed as a non-invasive measurement scheme of the 
plasma parameters -- the coupling strength and the screening length. This approach 
requires only input from a time-series of configuration snapshots and particle velocities with no recourse to additional 
information about the system. 
Our results should be directly applicable as a simple, yet reliable diagnostic method for a variety of experiments, including 
dusty plasmas, colloidal suspensions and ions in traps, and can be employed to facilitate 
comparisons between experiments, theory and simulations.  
\end{abstract}

\maketitle

\section{Introduction}
The Yukawa model~\cite{Hamaguchi1999} of interaction between charged particles is successfully used to describe a large variety of experimental systems. Among these are charged dust grains in complex plasma liquids~\cite{Piel2002,Chan2005,bonitz_ropp10} and 
colloidal suspensions of particles~\cite{Lowen2001}. Due to its simplicity and universality as well as its success in describing experimental phenomena, the Yukawa model has been the subject of extensive theoretical investigations. Static properties such as the triple 
point and phase transitions have been investigated in three dimensions~\cite{Robbins1988,Hamaguchi1996,Hamaguchi1997} 
and are of continuing interest in two dimensions where the question of the melting scenario is still a topic of some
controversy~\cite{Naidoo1994, Hartmann2005, Vaulina2006a, Hartmann2007a}. Furthermore, dynamic processes of Yukawa systems under investigation include transport properties such as diffusion~\cite{Ohta2000}, thermal conductivity 
and viscosity~\cite{Salin2002,Saigo2002,Donko2004a} in three dimensions. In two dimensions (2D), anomalous behaviour of the transport 
properties has been a subject of detailed investigations~\cite{Liu2007,Ott2008,Ott2009b,Donk'o2009}. 
The collective excitation spectra and waves are also a subject of ongoing interest, 
in three~\cite{Kalman2000} as well as in two~\cite{Kalman2004} dimensions 
and in quasi-2D systems~\cite{Donk'o2004} and systems under the influence of a magnetic field~\cite{Hou2009c,Bonitz2010, Ott2010}.

It is desirable to test these and other theoretical results with data obtained for example from dusty plasma and 
colloidal suspensions experiments or ions in traps~\cite{Dubin1999,Dantan2010}. These systems can be realized in a highly correlated yet liquid-like state, 
where particles are able to overcome the energy barriers of their local potential wells (``cages'') with a sufficiently high probability to destroy long-range order indicative of solid systems. Using advanced particle-recognition schemes, these experimental setups 
deliver reliable phase-space trajectories (position and velocities) of the individual particles in the correlated system, see, e.g., Ref.~\cite{bonitz_ropp10} and references therein. 

In order to compare theoretical results for Yukawa systems to experiments, one first has to make sure that the experiments are 
performed in conditions where the Yukawa model is applicable. This requires, in particular, that ion streaming and wake effects are negligible and 
the pair interaction potential is close to an isotropic one. These conditions are rather well known. For example, the particles 
should be located sufficiently far from the electrodes or from void regions. Once this has been established [and throughout this paper this will be assumed], it is crucial to infer from the experiment the two parameters governing 
the Yukawa model:
the Coulomb coupling parameter, $\Gamma$, and the inverse 
screening length $\kappa $ (see Section~\ref{sec:sim} for definitions). For $\kappa=0$, the Yukawa-model reduces to 
the well-known one-component-plasma (OCP) model. 

In the case of 2D dusty plasma setups, $\Gamma$ and $\kappa$ are often obtained by comparison of experimental 
wave dispersions with theoretical and simulation data~\cite{Nunomura2002,Sullivan2006}. Another approach is to 
excite sound waves by laser manipulation and measure the ratio of the sound speeds of compressional and shear waves which 
is a function of $\kappa$~\cite{Wang2001} or the transverse sound speed $C_t$ alone, which, however, is not very sensitive to variation 
in $\kappa$ ($\Delta C_t/C_t\lesssim 0.2$ for $0<\kappa<2$)~\cite{Nunomura2000,Peeters1987a}. 
The particle charge can alternatively be measured by observing the oscillation frequency of individual particles in the
sheath~\cite{Melzer1994,Trottenberg1995} or by theoretical models of dust charging~\cite{Shukla2001a}. A parameter combining $\Gamma$ and 
$\kappa$ non-uniquely into a single value can be obtained by observing the particle diffusion 
as a function of time alone~\cite{Vaulina2008a}. All these methods have a limited range of applicability and they are, in part, 
not easy to use. 

It is, therefore, desirable, to investigate alternative  or complementary methods which rely on easily accessible experimental data,
which is the goal of this paper. We first develop an effective coupling parameter $\Gamma^\ast$ for 2D Yukawa systems which can be 
used to compare structurally similar systems, i.e. systems with (almost) identical pair distributions, but different values of $\Gamma$ and $\kappa$. This enables the 
comparison of Yukawa systems \textit{across different values of $\kappa$}, and makes it possible to investigate the 
influence of the interaction range alone on the behaviour of a Yukawa system without simultaneously and inadvertently disrupting 
structural features. 

We then propose a \emph{reference data method} (RDM) for obtaining $(\Gamma, \kappa)$ simultaneously from equilibrium particle trajectories 
alone, requiring no further information about the system as input. 
Our approach towards the RDM consists of first obtaining reference data by molecular dynamics (MD) 
simulation and then condensing structural and dynamical data into classification numbers. Fitting the relations 
between these classification numbers and $(\Gamma, \kappa)$, we derive symbolic formulas to obtain the system parameters. These formulas 
for $\Gamma$ and $\kappa$ are fairly easy to use and provide an accuracy of the order of 10 percent which is sufficient for most experimental purposes. The RDM requires no external perturbation of the system. 

The remainder of this paper is structured as follows: In  Section~\ref{sec:sim}, we introduce our simulation technique. In Section~\ref{sec:eff_coup}, we derive the effective coupling parameter and compare it to other approaches. 
This coupling parameter is used in Section~\ref{sec:rdm} to derive the RDM scheme. We conclude in Section~\ref{sec:conclusion}. 

\section{Yukawa-model and Simulation technique}
\label{sec:sim}
The Yukawa model is described by a collection of particles whose pairwise interaction is given by the potential
\begin{equation}\label{eq:phi}
 \Phi(r) = \frac{q}{r} \,e^{-r/\lambda_D}\,.
\end{equation}
Here $r$ is the distance between two particles, $\lambda_D$ is the Debye screening length and $q$ is the charge 
of the particles. 
In thermal equilibrium, the Yukawa system is characterized by two dimensionless numbers: 
\begin{description}
 \item[i)] the Coulomb coupling parameter, 
$\Gamma=q^2/(a k_BT)$, and 
 \item[ii)] the dimensionless inverse screening length $\kappa=a/\lambda_D$, 
\end{description}
where (in 2D) 
$a=\left[n \pi \right]^{-1/2}$, $n$ is the areal number density of the particles, and $T$ is the temperature.  
Below we will use the (inverse of the) plasma frequency $\omega_p = \left[ nq^2 /(\varepsilon_0 m)\right]^{1/2}$ as the time scale and $a$ as the length scale. 

We obtain reference data for the Yukawa model by solving the resultant equations of motion for $N=4080\dots 16320$ particles interacting 
via the pair potential (\ref{eq:phi}) by
molecular dynamics simulations. The particles are uniform in mass $m$ and charge $q$ and are placed inside a 
rectangular simulation box subject to periodic boundary conditions. Forces are calculated by imposing a $\kappa$-dependent 
cut-off radius, when $\kappa>0$. For the Coulomb case of $\kappa=0$, we employ the Ewald summation technique~\cite{Ewald1921}. 
The system is put into equilibrium prior to data collection by continuous  rescaling of the particles velocities to the 
required temperature. During the data collection, the system evolves microcanonically. For details of the simulation, see Ref.~\cite{Ott2009}.

\section{Effective Coupling and Pair Distribution}
\label{sec:eff_coup}
In this section, we study the pair distribution function of  a 2D Yukawa system for different values of the coupling parameter 
and of the screening length. We discuss some previous work which introduced effective coupling parameters and propose an 
improved approach which is based on the shape of the pair distribution, more precisely, on the height of its first peak. 
In conclusion, we demonstrate that the proposed concept can be extended from the strongly coupled fluid regime also to the freezing point.

\begin{figure*}
\centering
\includegraphics[scale=0.515]{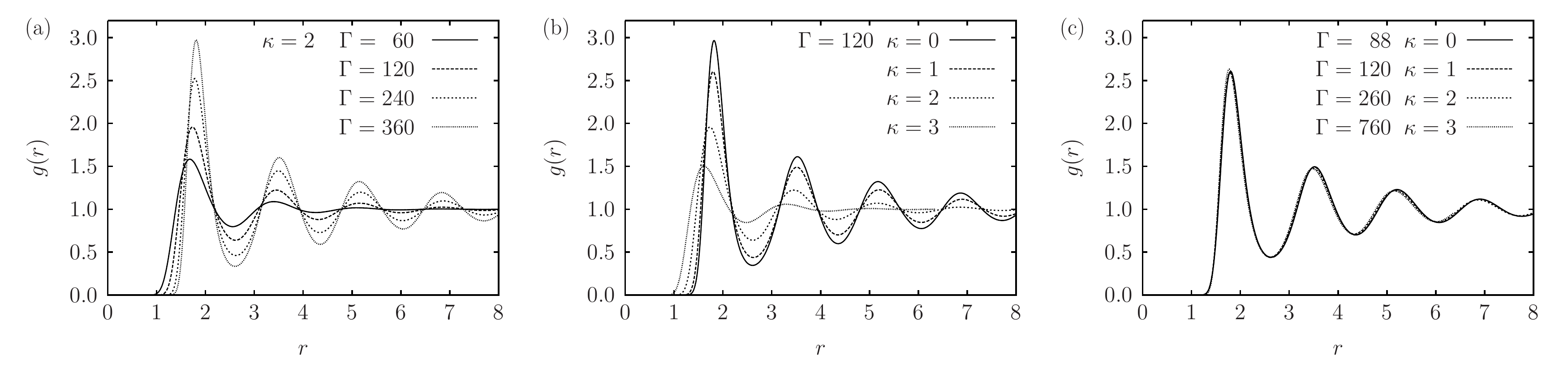}
\vspace{-0.5cm}
\caption{\label{fig:gofr} Pair distribution function of a strongly coupled Yukawa liquid. {\bf a)} at constant screening, $\kappa=2$, and different $\Gamma$, {\bf b)} at constant Coulomb coupling, $\Gamma=120$, and different $\kappa$ and {\bf c)} for four different combinations $\Gamma, \kappa$ which yield almost identical PDF, in particular a constant height of the first peak.}
\end{figure*}

\subsection{Pair distribution function}

The pair distribution function (PDF) is defined as 
\begin{equation}
 g(r) = \frac{A}{N^2} \left\langle \mathop{\sum\nolimits'}_{i,j=1}^N \delta(r-  r_{ij}) \right\rangle,
\end{equation}
where $r_{ij}=\vert \vec r_i - \vec r_j\vert$, $\langle . \rangle$ is a temporal average and the primed sum 
indicates the omission of the term $i=j$. It reflects the probability of finding two particles separated by $r$ in 
the system~\cite{Ott2009}.
The PDF of an ideal (non-interacting) system is unity, while correlations due to interactions give rise to regions of 
enhanced [$g(r)>1$] and reduced [$g(r)<1$] values of the PDF. 

Figure~\ref{fig:gofr}(a) shows the PDF in a Yukawa system ($\kappa=2$) 
for different values of $\Gamma$. All PDFs show a correlation hole at small $r$ due to particle repulsion and 
different degrees of modulations deriving from	 particle ordering, which become more pronounced as the coupling $\Gamma$ is increased. 
In the liquid phase, the order in the system is only short-ranged and the peaks of the PDF decay quickly with increasing $r$. 

The effect of the screening strength, $\kappa$, is analyzed in Fig.~\ref{fig:gofr}(b). With increased screening, the correlation hole 
narrows and the modulation of the PDF becomes less pronounced. For reasonably high coupling, Yukawa systems with different values 
of $\kappa$ possess a very similar PDF, if $\Gamma$ is 
chosen appropriately for each $\kappa$,  cf. Fig.~\ref{fig:gofr}(c). This behavior was observed by various authors, see, e.g., Ref.~\cite{Kalman2004} and references therein, and suggests that the structural properties of a Yukawa liquid do not depend separately 
on $\Gamma$ and $\kappa$ but rather on a single parameter {\em effective coupling parameter} which is a suitable combination of the two.

\subsection{Effective coupling parameters}
For the OCP, $\kappa=0$, the coupling parameter $\Gamma$ is equivalent to the ratio of the nominal 
nearest-neighbour interaction energy $q^2/a$ and the thermal energy $k_BT$. Clearly, for $\kappa>0$, such an interpretation 
is not possible, since the factor $\exp(-\kappa$) is missing from the former. Thus, the Coulomb coupling $\Gamma$ has no immediate 
physical meaning for systems with $\kappa>0$. 
The observation in Fig.~\ref{fig:gofr}(c) that Yukawa systems with different combinations of $(\Gamma, \kappa)$ show similar (static) properties has led to efforts to define a universal effective coupling parameter $\Gamma^\ast(\Gamma, \kappa)$ analogous to 
the OCP coupling parameter. Usually, $\Gamma^\ast$ is tailored to converge to $\Gamma$ as $\kappa$ goes to zero, 
so that for example the freezing of a Yukawa liquid occurs at the known freezing point of the OCP, $\Gamma^\ast=137$. 

Various authors have proposed different functional forms of $\Gamma^\ast(\Gamma,\kappa)$. The intuitive 
definition
\begin{equation}
 \Gamma^{\ast}_1(\Gamma,\kappa) = \Gamma \, e^{-\kappa}
\label{eq:gs1}\end{equation}
has been proposed by Ikezi and others, e.g.~\cite{Ikezi1986,Totsuji2001,Sorasio2003}. 
However, one problem with this approach is that
systems with the same $\Gamma^{\ast}_1$ do not exhibit many similarities and can be either in the 
liquid or in the solid phase, depending on $\kappa$~\cite{Vaulina2002a, Hartmann2005}. 

A different definition given by Vaulina and Khrapak~\cite{Vaulina2000},
\begin{equation}
 \Gamma^\ast_2(\Gamma,\kappa) = c\; \Gamma \, e^{-d\kappa} \left(1+d\kappa+\frac{(d\kappa)^2}{2} \right)\label{eq:gs2}\,,
\end{equation}
has been frequently used for three dimensional systems, where $c=1$~\cite{Vaulina2002a,Vaulina2004a,Fortov2003}. 
In two dimensions, the factor $c=1.5$ is sometimes used~\cite{Vaulina2006}. The factor $d$ in the above 
equations appears due to differences in the respective system of units and is $d=\pi^{1/2}$ in 2D and $d=(4\pi/3)^{1/3}$ in 3D. 
Equation~\eqref{eq:gs2} can be obtained by assuming that the coupling parameter is proportional to  $l^2\Phi''(a)/2$ 
at a characteristic distance~$l$ instead of being proportional to $\Phi(a)$. This expression is sensitive to the distance 
fluctuations of particles around a stationary state (characterized by the interparticle distance $a$) and should thus be more suitable to characterize the system behavior near 
the crystallization point than the absolute value of the potential itself.
A similar expression was used in Ref.~\cite{Bonitz2006} for finite three dimensional systems. 

For two dimensional systems, Hartman \textit{et al.}~\cite{Hartmann2005,Kalman2004,Hartmann2006} have defined an effective 
coupling parameter, 
\begin{equation}
\Gamma^\ast_3(\Gamma,\kappa) = f(\kappa)\Gamma, \label{eq:gs3}
\end{equation}
based on a constant amplitude of the first peak of the pair distribution function $g(r)$.  
The scaling function $f(\kappa)$ has been found by 
fitting numerical data to a polynomial in $\kappa$ up to fourth order which has the form
\begin{equation*}
 f(\kappa) = 1-0.388\kappa^2 + 0.138\kappa^3 - 0.0138\kappa^4.
\end{equation*}
The validity of the definition of $\Gamma^\ast_3$ is limited i) to the maximum value of $\kappa_\textrm{max}=3.0$ used 
to derive $f(\kappa)$~\cite{Ott2009a}, and ii) to a minimum value of $\Gamma^\ast_3\approx 40$. Below this latter value, the scaling 
is not exclusively dependent on $\kappa$ and thus no universal scaling function $f(\kappa)$ can be given, as will be shown below. 

The ratio of the coupling parameter and the coupling parameter $\Gamma_c(\kappa)$ at the crystallization transition temperature has also 
been employed to quantify the actual physical coupling of the system~\cite{Liu2006, Ott2009a,Ott2009c}:
\begin{equation}
\Gamma^\ast_4(\Gamma,\kappa) = \frac{\Gamma_c(0)}{\Gamma_c(\kappa)}\Gamma.
\label{eq:gs4}
\end{equation}
This definition correctly captures the main structural properties of a Yukawa system but its validity range remains unclear.

Our goal in this paper is to find an effective coupling parameter $\Gamma^\ast$ which is valid (at least) in the entire liquid phase. 
It should correctly reflect the structural properties of the system which are fully contained in the pair distribution function.
For strongly coupled systems, the first peak amplitude of the PDF already 
characterizes the complete PDF, except for very high values of $r$ (see discussion in Ref.~\cite{Hartmann2005}). For less 
strongly coupled liquids, the size of the correlation hole and the position of the subsequent peaks of the PDF differ for systems 
with different values of $(\Gamma, \kappa)$, even if the first peak height of the PDF is identical. 
Using the first peak height to characterize the structural properties of the system is, of course, only one of several 
possible choices but
offers numerous advantages; it is, for example, clearly defined, dimensionless and easily accessible in simulations and 
experiments. 

We, therefore, employ a similar approach as Hartmann~\textit{et al.} by finding a 
scaling function $w$ which results in a constancy of the first peak amplitude of the PDF. 
However, to remove the limitations of their formula we allow for a $\Gamma$-dependence of the scaling function, 
\begin{equation}
\Gamma^\ast(\Gamma,\kappa) = w(\Gamma,\kappa)\Gamma \label{eq:gstarn}\,,
\end{equation}
and extend our definition to $0.0<\kappa\leq 5.0$. In addition, we are interested in the inverse relation
\begin{equation}
\Gamma(\Gamma^\ast,\kappa) = w^\ast(\Gamma^\ast,\kappa)\Gamma^\ast \label{eq:gstarnr}. 
\end{equation}
\begin{figure}
\includegraphics[scale=0.65]{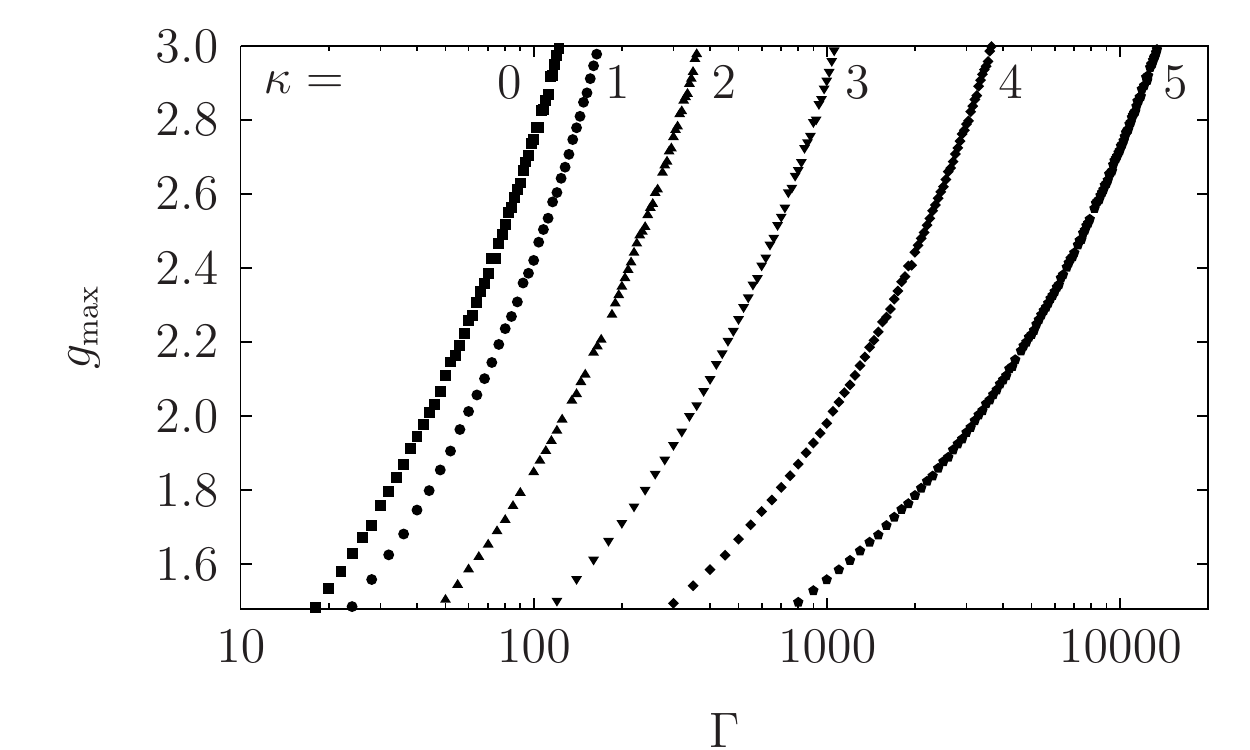}~~~~~~~~
\caption{\label{fig:peakheight} The height $g_\textrm{max}$  of the first peak of the PDF as a function of~$\Gamma$. 
}
\end{figure}
As a first step, we obtain -- by first-principle MD simulations -- the reference data for $\kappa=0\dots 5$ in steps of $\Delta\kappa=0.5$ and 60-90 values of $\Gamma$ per $\kappa$. From these data, we 
read off the PDF peak height $g_\textrm{max}$ as a function of $\Gamma$ and $\kappa$. 
Some of the resulting data are shown in Fig.~\ref{fig:peakheight}. We chose an interval
$g_\textrm{max}\in (1.5,3.0)$ to carry out the subsequent analysis.

\begin{figure}
\includegraphics[scale=0.65]{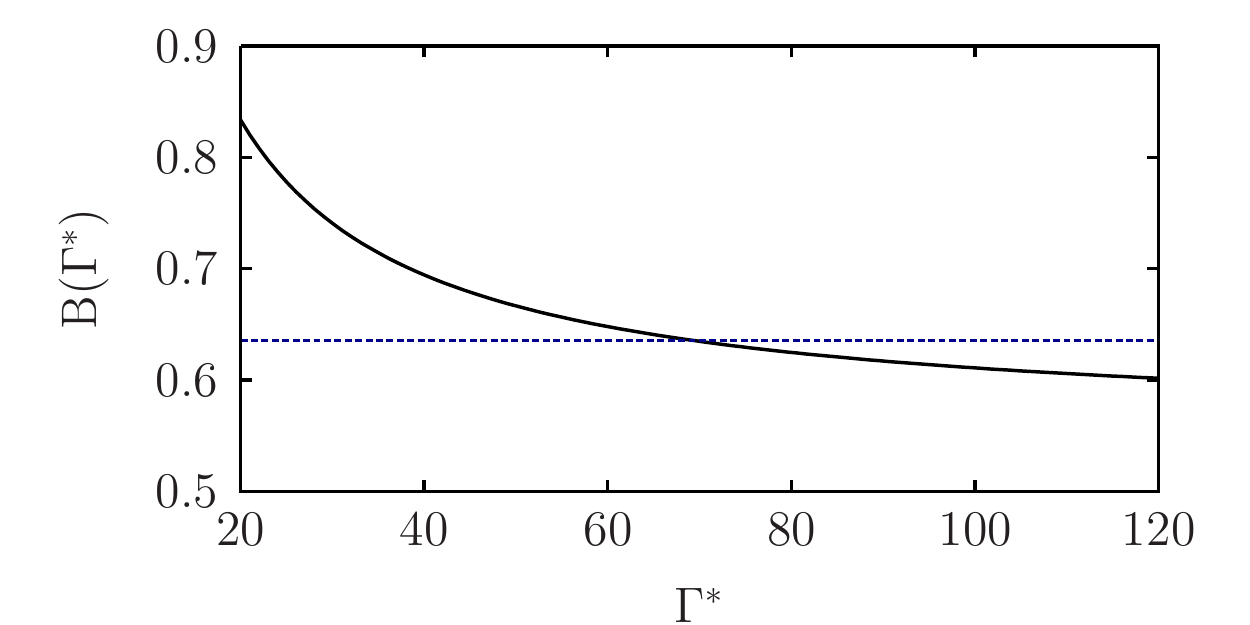}~~~~~~~~
\caption{\label{fig:eqb} $B(\Gamma^\ast)$ according to Eq.~\eqref{eq:b}. The horizontal 
line is the constant approximation used in Eq.~\eqref{eq:simplew} (see text). }
\end{figure}

The ratio $g_\textrm{max}(\Gamma,\kappa)$ normalized by $g_\textrm{max}(\Gamma,0)$ 
is equivalent to  $w^\ast(\Gamma^\ast,\kappa)$. Employing symbolic regression~\cite{Schmidt2009}, 
we find that the following functional form provides excellent agreement with the simulation data:
\begin{eqnarray}
w^\ast(\Gamma^\ast,\kappa) &=& \exp\left(\frac{\kappa^2}{A + B(\Gamma^\ast)\kappa}\right)\label{eq:gs}\\
B(\Gamma^\ast) &=& B_1 + B_2/\Gamma^\ast\label{eq:b}\,,
\end{eqnarray}
where $A=2.37221$, $B_1=0.55515$, and $B_2=5.56585$. 
For reasonably high coupling, $\Gamma^\ast \gtrsim 40$, the function $B(\Gamma^\ast)$ is only weakly dependent on $\Gamma^\ast$ (cf. Fig.~\ref{fig:eqb}) and 
is thus well approximated by a constant. This leads to a simpler approximate 
effective coupling which depends on $\kappa$ only, 
in line with the earlier definitions \eqref{eq:gs1}-\eqref{eq:gs4}, 
\begin{eqnarray}
 \Gamma &\approx& w^\ast(\kappa)\Gamma^\ast, \qquad \Gamma^\ast\gtrsim 40 \label{eq:simplew}
\end{eqnarray}
where $w^\ast(\kappa)=w^\ast(\Gamma,\kappa)$, as defined above but with 
 $A=2.33425$, $B_1=0.63529$, and $B_2=0$, resulting from the described averaging procedure.

This relation is readily inverted, yielding
\begin{eqnarray}
\Gamma^\ast &\approx& w(\kappa)\Gamma\,,
\end{eqnarray}
where $w(\kappa)=1/w^\ast(\kappa)$. While the general solution of Eqs.~\eqref{eq:gstarnr}, \eqref{eq:gs} for $\Gamma^\ast$ is quite challenging and can only be achieved numerically,
we have found that the following symbolic expression can be used as an approximation, in lieu of the numerical inversion:
\begin{eqnarray}
 w(\Gamma,\kappa) &=& C(\kappa) + \frac{D(\kappa)}{\Gamma}\\
 C(\kappa) &=& (C_1+C_2\kappa^2)^{C_3{\kappa^2}}\\
 D(\kappa) &=& \frac{\kappa^3}{(D_1\kappa^2+D_2)}
\end{eqnarray}
Here $C_1=0.0063503$, $C_2=0.00209988$,$C_3=0.0668851$ and $D_1=0.309364$, $D_2=1.749$.

\subsection{Analysis of formula~\eqref{eq:simplew}}

\begin{figure}
\includegraphics[scale=0.65]{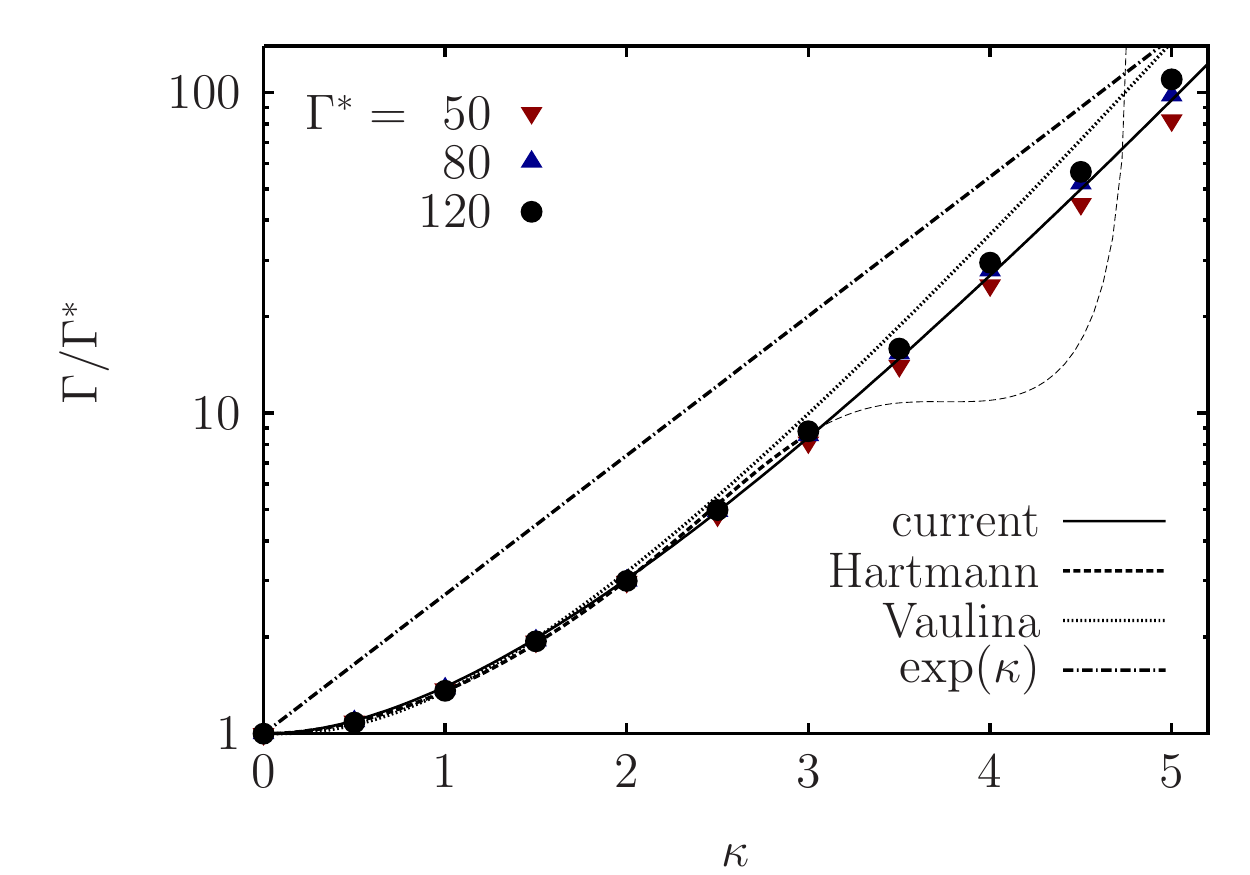}~~~~~~~~
\caption{\label{fig:comp1} The ratio $\Gamma/\Gamma^\ast$ as a function of $\kappa$. Symbols denote 
data taken from Fig.~\ref{fig:peakheight} and the full line corresponds to the scaling function 
according to the fit formula~\eqref{eq:simplew} of this paper. The expression~\eqref{eq:gs3} of Hartmann~\textit{et al.}, 
the formula~\eqref{eq:gs2} of Vaulina~\textit{et al.} ($c=1$), and the expression~\eqref{eq:gs1} by Ikezi are shown 
for comparison. 
The thin dotted line is the continuation of~\eqref{eq:gs3} beyond its scope of validity. }
\end{figure}

\begin{figure}
\includegraphics[scale=0.65]{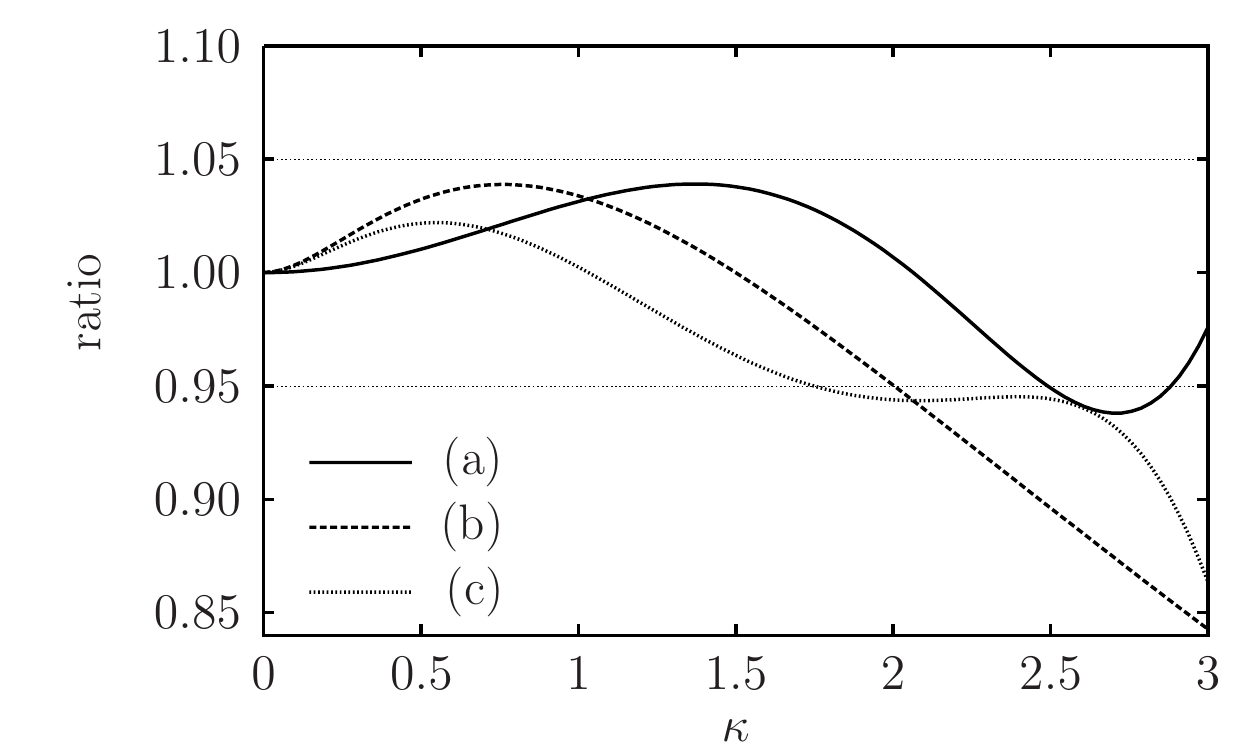}~~~~~~~~
\caption{\label{fig:comp3} $\kappa$-dependence of the ratios of pairs of different scaling functions. 
(a) Our formula~\eqref{eq:simplew} to formula~\eqref{eq:gs3} [Hartmann~\textit{et al.}],
(b) Formula~\eqref{eq:simplew} to formula~\eqref{eq:gs2} [Vaulina~\textit{et al.}],
(c) Formula~\eqref{eq:gs3} to formula~\eqref{eq:gs2}. 
The ratio (c) continues to decay monotonically beyond $\kappa=3$; the horizontal lines 
mark a deviation of $\pm5\%$.}
\end{figure}

\begin{figure}
\includegraphics[scale=0.65]{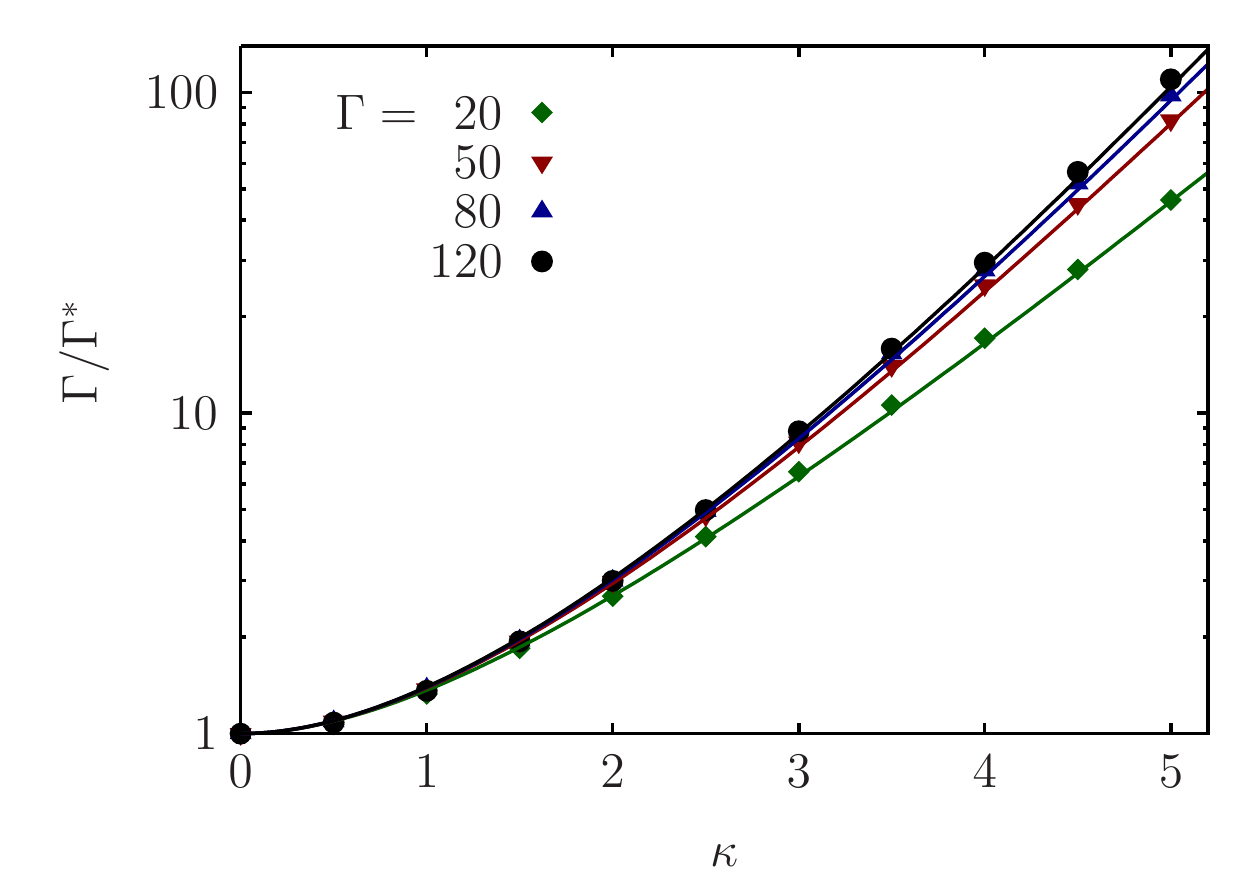}~~~~~~~~
\caption{\label{fig:comp2} The ratio $\Gamma/\Gamma^\ast$ as a function of $\kappa$. Symbols denote 
data taken from Fig.~\ref{fig:peakheight} and the full lines correspond to the scaling function 
according to our full model~\eqref{eq:gs}.  }
\end{figure}

\begin{figure}
\includegraphics[scale=0.65]{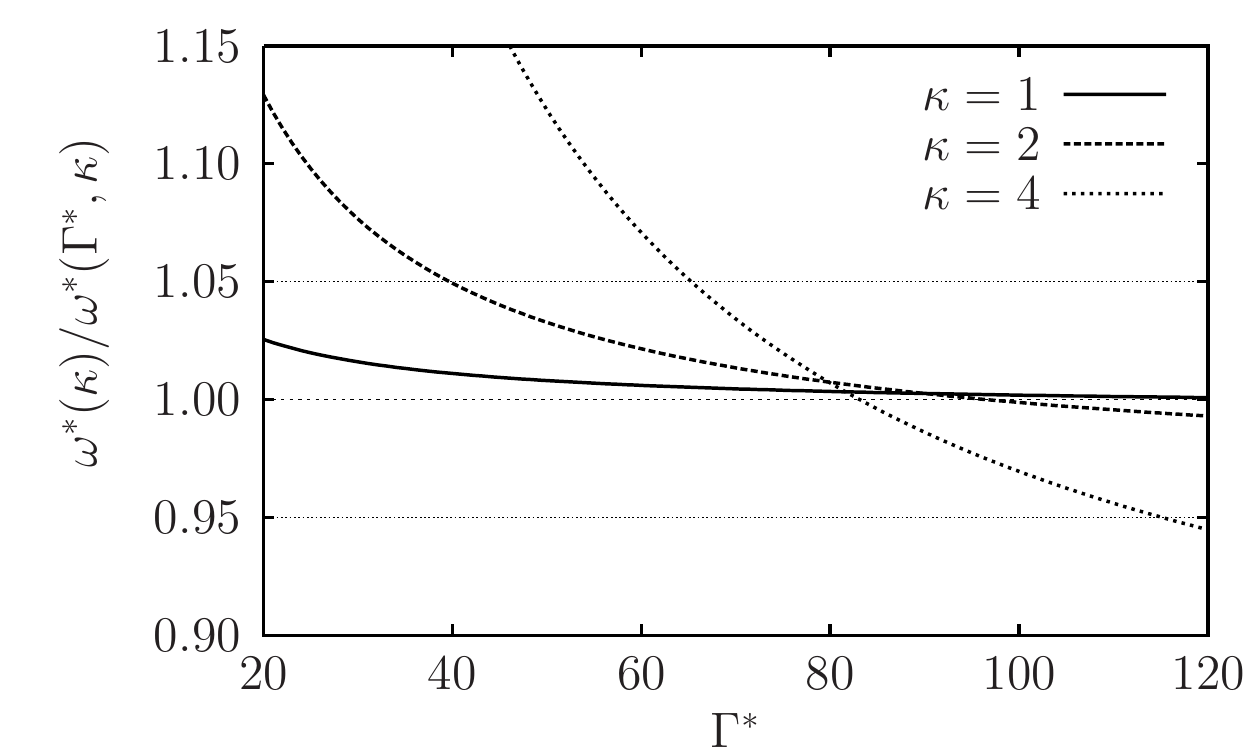}~~~~~~~~
\caption{\label{fig:comp4} The ratio of the $\Gamma^\ast$-independent scaling function $w^\ast(\kappa)$  
to the full expression $w^\ast(\Gamma^\ast, \kappa)$ for three values of $\kappa$. The horizontal lines 
mark a deviation of $\pm5\%$. }
\end{figure}

Let us first compare the $\Gamma^\ast$-independent definition~\eqref{eq:simplew}  with the alternative 
scaling functions which were proposed earlier, cf. Eqs.~(\ref{eq:gs1})--(\ref{eq:gs3}). They are shown in Fig.~\ref{fig:comp1} together with 
the actual ratio of the peak heights obtained from the MD simulations which are indicated by symbols. 
Obviously, the simulation data for different values of $\Gamma^\ast$ 
do not fall on a universal curve; therefore, the scaling cannot be a universal function of $\kappa$. How well a particular 
scaling function fits the data is dependent on the interval of $\Gamma^\ast$ which it aims to describe. We also note 
that the exponential and Vaulina's definition are not, unlike Hartmann's and ours, specifically aimed at describing 
the structural order and are thus not expected to coincide with the data points of Fig.~\ref{fig:comp1}.

Apart from the exponential scaling, the curves nonetheless show a nearly universal behaviour. 
Below $\kappa\approx 2.5$, there are essentially only small differences. Vaulina's scaling is slightly higher 
than the other two and fits the structural data somewhat worse. Differences between Hartmann's et al. and our scaling are 
small and likely due to the $\Gamma^\ast$ interval of the respective validity, as discussed above. Note however, that our 
scaling $\omega^\ast(\kappa)$ contains only two numerical fit parameters as opposed to Hartmann's three. 

Above $\kappa=3$, Hartmann's scaling is not valid any more and its numerical extrapolation, shown by the thin dotted line in Fig.~\ref{fig:comp1} clearly gives 
unphysical results. Vaulina's and our scaling differ appreciably in this range of $\kappa$. 
To quantify the respective deviations, various ratios of pairs of scaling functions 
are shown as a function of $\kappa$ in Fig.~\ref{fig:comp3}. It can be seen that the difference between Hartmann's scaling and 
ours only slightly exceed 5\% for $\kappa\leq 3$ while differences between Vaulina's and our scaling 
quickly grow with increase of $\kappa$ and exceed 15\% for $\kappa>3$. 

\subsection{Analysis of formula~\eqref{eq:gstarnr}}

We derived the $\Gamma^\ast$-independent scaling by averaging the full model expression over its $\Gamma^\ast$-dependence for $\Gamma^\ast>40$. 
As can be seen in Fig.~\ref{fig:comp1}, this procedure gives best agreement for  $\Gamma^\ast\approx 80$ but it 
fails for smaller values of $\Gamma^\ast$.
Thus, for systems with lower coupling, the full scaling function~\eqref{eq:gs} should be used instead. In Fig.~\ref{fig:comp2}, 
this scaling function is shown for four values of $\Gamma^\ast$ covering all but the most extreme states of liquid systems. 
The excellent agreement is evident, even though only three free parameters enter into the scaling function. 

Finally, we directly compare our two models: the approximate $\Gamma^\ast$--independent scaling function, Eq.~(\ref{eq:simplew}), and the full $\Gamma^\ast$--dependent counterpart, Eq.~(\ref{eq:gstarnr}). Figure~\ref{fig:comp4} shows, for three values of $\kappa$, their ratio as a function of $\Gamma^\ast$. 
As expected, deviations are smallest for $\Gamma^\ast\approx 80\dots 100$ and increase with increasing distance from this range. 
The relative deviations are smaller for small $\kappa$ and are less than 3\% for $\kappa=1$ and $\Gamma^\ast \in [20,120]$. 
For higher $\kappa$, the deviations increase substantially as the spread of the ratio $\Gamma/\Gamma^\ast$ becomes larger 
(cf. symbols in Fig.~\ref{fig:comp1}). 

The effective coupling parameter introduced in this section can serve a number of purposes. It can be used to 
compare systems with different interactions but identical structural properties by varying the screening from $\kappa=0$ (long-ranged 
interaction) to $\kappa=5$ (short-ranged interaction). Such a line of research promises to give insight into the 
dependence of physical properties on the range of interparticle interaction. Structural properties such as the 
crystallization temperature can be obtained by taking the well-known results for the OCP and transferring 
them, via $\Gamma^\ast$, to Yukawa systems. 
Since $\Gamma(\Gamma^\ast,\kappa)$ grows monotonically with $\kappa$, it is also possible to study dynamical properties of systems with 
identical structure but different thermal excitation (recall that $\Gamma\propto T^{-1}$). This allows one to 
separate the influence of thermal excitation and structural changes on dynamical properties, which are otherwise changed 
simultaneously when $T$ is varied~\cite{Ott2010}. 

\subsection{Pair distribution and crystallization}
While our present analysis focuses on strongly coupled Yukawa fluids it is, nevertheless, tempting to take a brief look 
at larger couplings where the system freezes. The change of the pair distribution function (and of related functions, such as 
the static structure factor) at the liquid--solid transition 
has been studied in great detail by many authors~\cite{Hansen1969,Schiffer2002,Sastry2003, Boning2008}. 
Yet we are not aware of a systematic analysis of the height of the first peak 
of $g(r)$ at the phase transition. We have, therefore, performed extensive additional MD simulations for $\Gamma$ and $\kappa$ values 
in the vicinity of the freezing transition. 

\begin{figure}
\includegraphics[scale=0.65]{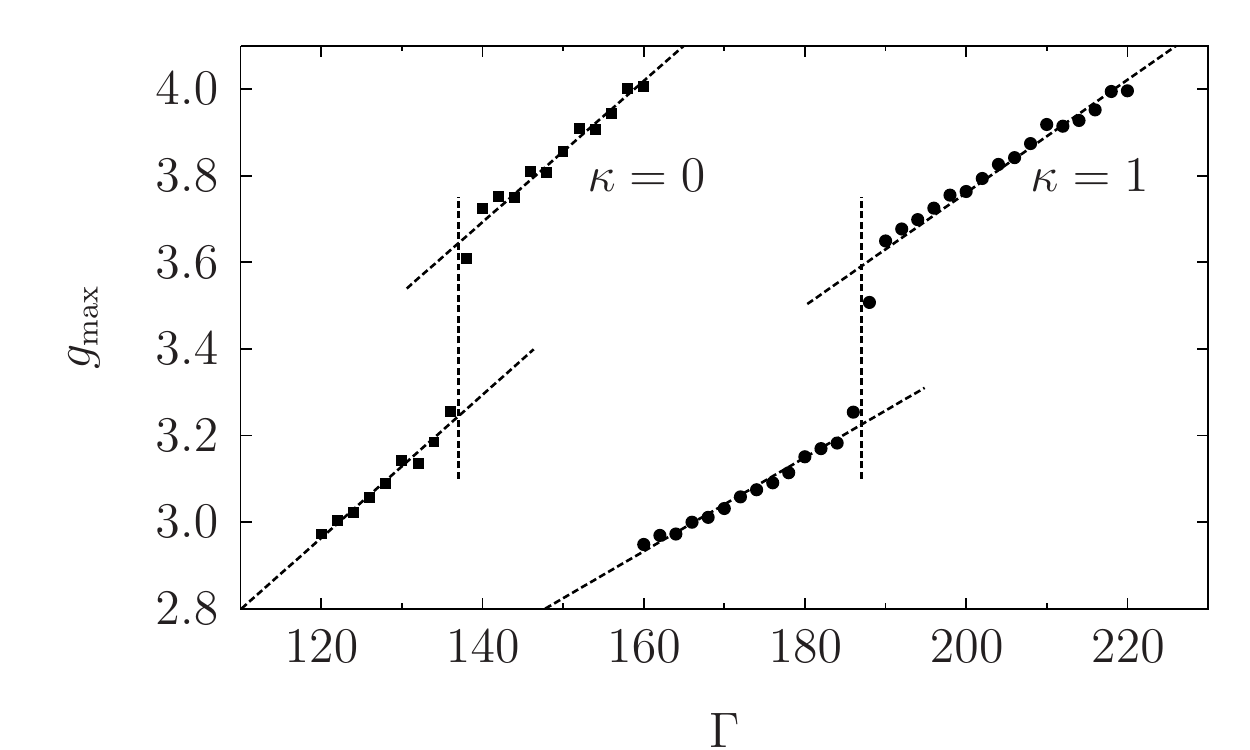}~~~~~~~~
\caption{\label{fig:peakheight_melt} The height $g_\textrm{max}$  of the first peak of the PDF as a function of~$\Gamma$ in the 
vicinity of the liquid-solid transition for a Coulomb system and a Yukawa system with $\kappa=1$. The vertical lines mark the 
crystallization points, $\Gamma_c=137$ and $\Gamma_c=187$, respectively. 
}
\end{figure}

Some representative results for the first peak height $g_{\rm max}$ of the PDF are shown in Fig.~\ref{fig:peakheight_melt} and extend the data of Fig.~\ref{fig:peakheight} to larger couplings. The first observation is that $g_{\rm max}$ changes discontinuously. Comparing to the known freezing points of Yukawa and Coulomb systems we observe that the position of this jumps exactly agrees with these values. We thus conclude that the height of the  first peak of $g(r)$ may serve as valuable indicator of the liquid-solid transition in one-component Yukawa systems. This value is fairly easily detected in an experiment and does not require long measurements to achieve convergence. Preliminary results indicate that this behavior is observed also for other types of interactions and for phase transitions in quantum systems~\cite{Bonitz2008,Filinov2001,Filinov2008} as well.
%

\section{The reference data method}
\label{sec:rdm}

In the previous section, we have introduced an effective coupling parameter $\Gamma^\ast$  which is based on the constancy 
of the first peak of the PDF, i.e. on the universality of short-range structural features. However, even if two systems are practically indistinguishable in their structural properties, and thus have identical $\Gamma^\ast$, in general, these systems will exhibit differences in their dynamical behaviour. We can reformulate this statement in the following way: If we could measure structural and dynamical properties, this information should be sufficient to distinguish Yukawa systems with respect to both their coupling and screening parameter. 

We will now exploit this fact 
to derive a practical method which allows to calculate both the values of $\Gamma$ and $\kappa$ simultaneously from a \textit{time-ordered 
series of particle snapshots alone.} No assumptions about the time scale of the snapshots are being made, 
except that they are evenly spaced in time (a restraint easily accomplished in experiments, but easily dropped if needed) 
and of sufficiently rapid succession to resolve single-particle oscillations. This allows for computing the instantaneous 
particle velocities. 

As a first step, we obtain a direct mapping between the peak height $g_\textrm{max}$ of the PDF and the 
effective coupling parameter introduced earlier.  We find the following simple relation:
\begin{equation}
 \Gamma^\ast(g_\textrm{max}) = 19.2581\,g_\textrm{max}^2 - 16.6829\,g_\textrm{max}, \label{eq:gs_gm}
\end{equation}
which allows one to calculate $\Gamma^\ast$ without prior knowledge of $\kappa$ and $\Gamma$ [and so far without  
knowledge of the dynamical properties] of the system.  Simply by calculating $g(r)$ in a very 
limited range of $r$, as can be done from a couple of system snapshots, one can obtain the effective 
coupling which governs the system's static and structural behaviour. 

What remains to be achieved is to extract $\Gamma$ and $\kappa$ from the result (\ref{eq:gs_gm}) which does require additional 
information. A suitable supplementary quantity is the velocity autocorrelation function which we discuss in the following.

\subsection{Velocity autocorrelation function}

\begin{figure}
\includegraphics[scale=0.65]{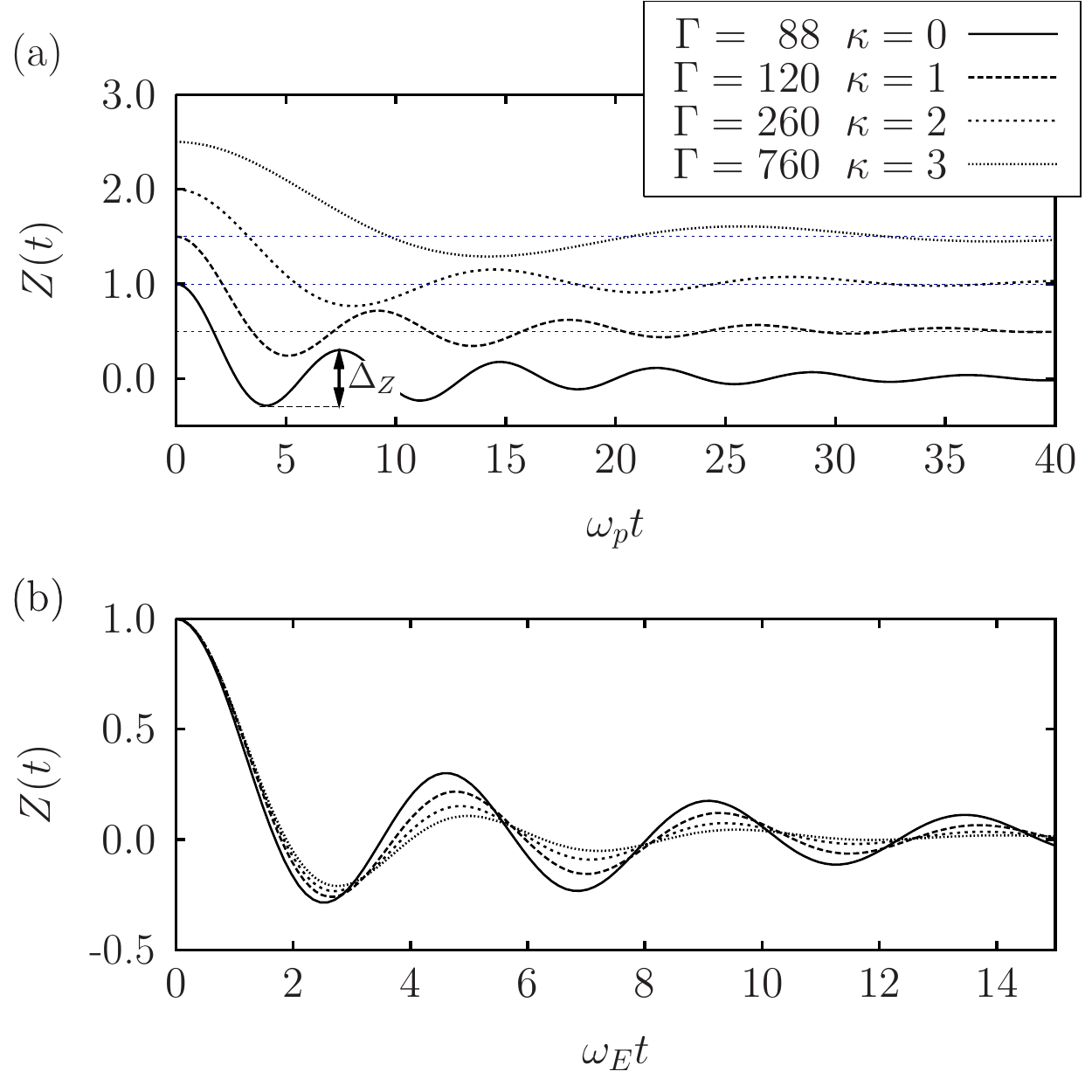}~~~~~~~~
\caption{\label{fig:vacf} Time dependence of the velocity autocorrelation function $Z(t)$ for systems with identical short-range 
order, i.e. identical $\Gamma^\ast$, cf. Fig.~\ref{fig:gofr}(c). {\bf (a):} Time argument normalized to the nominal plasma frequency. {\bf (b):} Time normalized to the Einstein frequency. Curves in (a) are shifted for clarity [$Z(0)=1$ in all cases].  
The arrow in (a) indicates the definition of $\Delta_Z$ (see text).   }
\end{figure}

The (normalized) velocity autocorrelation function (VACF) $Z(t)$ is defined as
\begin{equation}
 Z(t) = \frac{\langle \vec v(t)\cdot\vec v(0) \rangle}{\langle\vert \vec v(0)\vert^2\rangle}\,, 
\end{equation}
where 
$\langle .\rangle$ is an ensemble average which is performed by averaging over all particles and $\vec v$ denotes 
the particle velocity. 

The VACF is a fundamental measure of the dynamics of an ensemble of particles and is intimately connected, e.g., to the 
diffusion process. In reasonably highly coupled liquids, at any time instant, a fraction of particles oscillate in 
their local potential cages before diffusing farther. These periodic trajectories manifest themselves in an oscillatory VACF, 
which decays with time because of anisotropies of the cages and the increasing number of particles which leave their 
potential cage.

Figure~\ref{fig:vacf}(a) shows the VACF for systems with identical effective coupling $\Gamma^\ast$ but different $\kappa$. 
The oscillation period of the VACF curves differs appreciably across different values of $\kappa$. 
However, when replotted as a function of time $t$  multiplied by the Einstein frequency $\omega_E$, cf.~Fig.~\ref{fig:vacf}(b), the VACF oscillation 
frequency becomes very similar~\cite{Hartmann2005}. This is not surprising, since $\omega_E$ is defined as the (average) oscillation 
frequency of a single particle in the frozen environment of all other particles and thus takes into account the range of the pair  interaction. 

Our goal is to use the dynamical information contained in the VACF to distinguish between systems with the same short-range 
structural order (i.e., systems with the same $\Gamma^\ast$) but different $\kappa$. 
If one would have knowledge of the nominal plasma frequency $\omega_p$
of the system, this would be easily accomplished by using the $\kappa$-dependence of $\omega_E$ or, equivalently, 
of the VACF oscillations. However, 
$\omega_p$ is usually not known and cannot be calculated from the trajectory snapshots alone. Therefore, 
we must choose a criterion which is independent of the time scale. While the periodicity of the VACF can be mapped on a 
single curve for different $\kappa$, Fig.~\ref{fig:vacf}(b) shows that, nevertheless, there are clear $\kappa$-dependent differences 
in the amplitude of the oscillations. This suggests to use, e.g., the amplitude difference $\Delta_Z$ between 
the first minimum and the second maximum of the VACF as a parameter suitable to distinguish between different screening parameters, see Fig.~\ref{fig:vacf}(a). 

\begin{figure}
\includegraphics[scale=0.6]{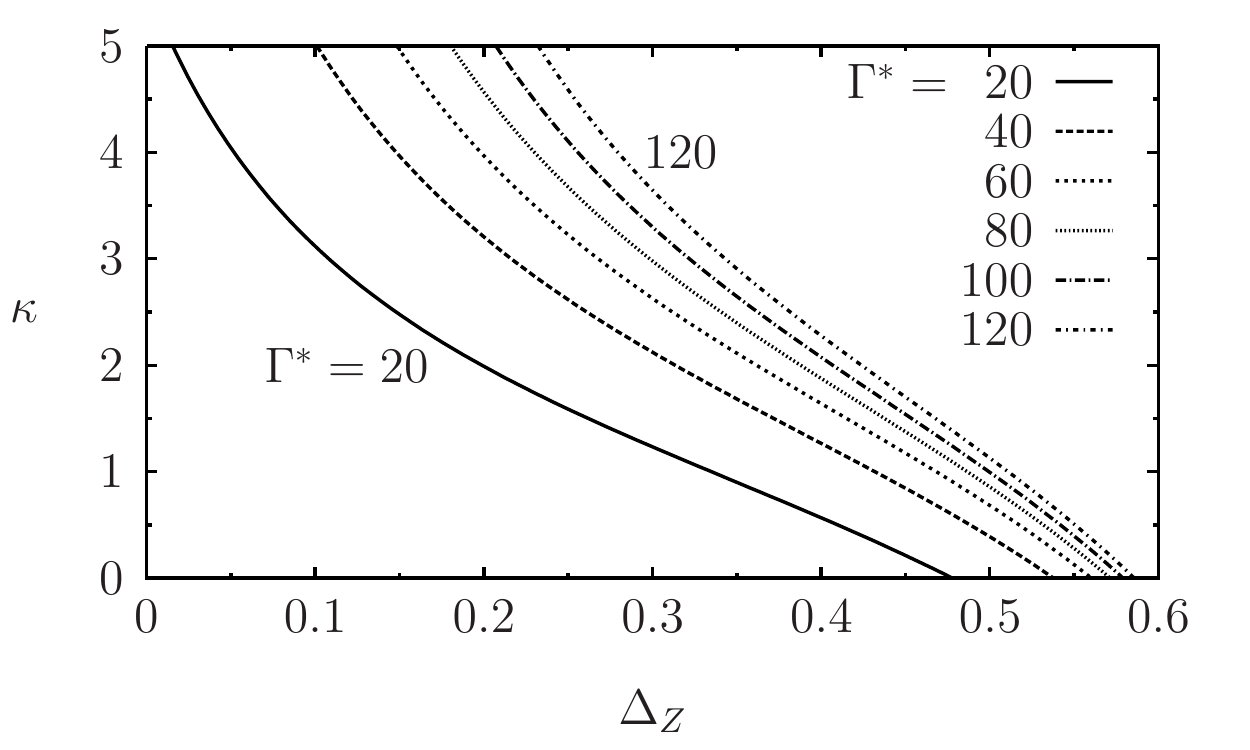}
\caption{\label{fig:k_of_z} The screening parameter $\kappa$ as a function of $\Delta_Z$ (Eq.~\eqref{eq:kappa_delta})
for different values of $\Gamma^\ast$. }
\end{figure}

Using our reference data as before, it is possible to derive an explicit symbolic approximation for the screening parameter $\kappa$ 
as a function of the two ``experimental'' parameters $\Delta_Z$ and $\Gamma^\ast(g_{\rm max})$:
\begin{eqnarray}
 \kappa(\Delta_Z, \Gamma^\ast) &=& \frac{A_\Delta(\Gamma^\ast)}{B_\Delta(\Gamma^\ast)+\Delta_Z} + \frac{\Delta_Z}{\Delta_Z-C_\Delta}
 \label{eq:kappa_delta}\\
A_\Delta(\Gamma^\ast) &=& 1.28486 - {\Gamma^\ast}^{f_A(\Gamma^\ast)}\nonumber\\
f_A(\Gamma^\ast) &=& {0.00171645-0.011499\,\Gamma^\ast} \nonumber\\
B_\Delta(\Gamma^\ast) &=& 0.166741-0.0013331\,\Gamma^\ast\nonumber\\
C_\Delta &=& 0.856117 \nonumber\,.
\end{eqnarray}
This result for $\kappa(\Delta_Z, \Gamma^\ast)$ is shown in Fig.~\ref{fig:k_of_z} as a function of $\Delta_Z$, for various values of $\Gamma^\ast$.
The somewhat convoluted form of Eq.~\eqref{eq:kappa_delta} results from a strong sensitivity of the functional form of the curves on 
their parameters. 
Figure~\ref{fig:k_of_z} also demonstrates that $\Delta_Z$ is a viable choice to characterize $Z(t)$ as a function 
of $\Gamma^\ast$ and $\kappa$: The curves do not cross and proceed monotonically both for fixed $\Gamma^\ast$ and 
fixed $\kappa$. In addition, the values of $\Delta_Z$ span a sufficiently broad interval to make 
an accurate separation possible. 

\subsection{Applying the RDM}
The proposed reference data method is now a straightforward application of the results derived before. It can be written down as follows:
\begin{enumerate}
 \item From the particle snapshots, calculate $g(r)$ and $Z(t)$ and read off $g_\textrm{max}$ and $\Delta_Z$.
 \item Calculate $\Gamma^\ast$ using Eq.~\eqref{eq:gs_gm} and $g_\textrm{max}$.
 \item Use Eq.~\eqref{eq:kappa_delta} to calculate $\kappa$ from $\Gamma^\ast$ and $\Delta_Z$. 
\item Calculate $\Gamma$ using Eqs.~(\ref{eq:gstarnr}) and~\eqref{eq:gs} or  Eq.~\eqref{eq:simplew}.\\

\end{enumerate}
To assess the accuracy of the RDM, we use it to calculate $\Gamma$ and $\kappa$ from the $g_\textrm{max}$ and $\Delta_Z$ values 
of our MD simulations. Some typical examples are shown in Tab.~\ref{tab:examples}.
The full results (Tab.~\ref{tab:acc}) indicate an average absolute error of $\Delta\kappa\lesssim 0.15$ 
(recall that $\kappa$ is given in  dimensionless units) and a relative error 
$\Delta\Gamma/\Gamma\lesssim 0.15$ for the experimentally most relevant values of $\kappa\leq 3$.

\begin{table}[tp]
 \centering
\renewcommand{\arraystretch}{1.1}
\begin{tabular}{cccc}
\hline\hline
\phantom{xxx}$\kappa_\textrm{sim}\phantom{xxx}$ &\phantom{xxx}$\Gamma_\textrm{sim}$ \phantom{xxx} & \phantom{xxx}$\kappa$\phantom{xxx} & \phantom{xxx}$\Gamma\phantom{xxx}$  \\\hline
$0$ & 60 & -0.09 & 65.93 \\ 
$1$ & 100 & 1.12 & 109.0 \\ 
$2$ & 300 & 2.07 & 323.8 \\ 
$3$ & 300 & 3.01 & 297.8 \\ 
$4$ & 1000 & 4.07 & 1062 \\ 
$5$ & 3000 & 5.06 & 3246 \\ 

\hline
\end{tabular}
\caption{\label{tab:examples} Examples for the application of the RDM. 
The actual parameters of the simulation carry the subscript ``$\textrm{sim}$'' (first and second column), the results 
from the RDM are in the third and fourth columns} 
\end{table}

\section{Conclusion}
\label{sec:conclusion}
In this work, we have derived an effective coupling parameter $\Gamma^{\ast}$ for strongly correlated 2D Yukawa systems which is based on the constancy of the first peak of the PDF. For high coupling, $\Gamma^\ast\gtrsim 40$, a simple one-parameter relation holds, 
$\Gamma^{\ast}/\Gamma=w(\kappa)$. This expression was tested against previous definitions of an effective coupling parameter and shown to give similar ($\kappa\leq 3$) or even substantially improved ($\kappa>3$) results in terms of short-range structure. 

In the low coupling regime, however, no such simple form exists, and we showed how to define an appropriate coupling parameter 
which is of the form $\Gamma^\ast/\Gamma=w(\kappa,\Gamma)$. It again allows for an adequate representation of the structural features of Yukawa systems at these couplings. Only two (three) free parameters enter the proposed functional form for high (low) coupling. 

In the second part, we have outlined a reference data method to calculate the values of $\Gamma$ and $\kappa$ of a Yukawa 
system from a time series of configuration snapshots. Using the dynamical information contained in the first oscillation cycle 
of the VACF, we were able to devise a method which does not depend on additional physical information such as temperature, mass, 
charge or density, to achieve this goal. We tested this method and found that it gives reliable results with an error 
not exceeding $10\dots 15\%$ for majority of relevant situations. It should be directly applicable to a number of systems of charged particles, in particular dusty plasmas. 

We mention two caveats one should be aware of when applying the results of this work: 1) the effective coupling parameter 
and the RDM were derived for PDF peak heights in the range of $g_\textrm{max}=1.5\dots 3.0$  which corresponds to $\Gamma^\ast\approx 18\dots 124$), and 
are thus restricted to this range; 2) the effect of friction on the dynamics of the system has not been considered. This is not 
of importance for $\Gamma^\ast$ but more so for the RDM. Future work will aim to relax these restraints. 

Furthermore, we have shown that the peak height of $g(r)$ is also sensitive to the liquid-solid transition and may serve as a simply accessible first estimate. Finally, while our results were derived for spatially homogeneous macroscopic 2D Yukawa systems we expect 
that they can be straightforwardly extended to systems in an external trap such as atoms, molecules, ions or dusty plasmas. This is based 
on the observation \cite{wrighton_pre09,wrighton_cpp10} that the shell structure of these systems is accurately described within a theory which uses the pair correlations of the associated homogeneous system. This question is the subject of ongoing work.

\begin{table}[!tp]
 \centering
\renewcommand{\arraystretch}{1.1}
\begin{tabular}{ccc}
\hline\hline
\phantom{xxxxx}$\kappa_\textrm{sim}\phantom{xxx}$ &\phantom{xxxxx} $\langle\vert \Delta\kappa\vert\rangle$\phantom{xxxxx} & $\langle\vert\Delta\Gamma/\Gamma_\textrm{sim}\vert\rangle$ \\\hline
$0$ & 0.10 & 0.12 \\ 
$1$ & 0.15 & 0.10 \\ 
$2$ & 0.09 & 0.03 \\ 
$3$ & 0.05 & 0.12 \\ 
$0\dots 5$& 0.12 & 0.12\\
\hline
\end{tabular}
\caption{\label{tab:acc} Accuracy of the RDM for different values of $\kappa_{\rm sim}$, averaged over $\Gamma^\ast$. The second (third) column 
contains the absolute (relative) deviations of the RDM results from the original simulation parameters.} 
\end{table}

\begin{acknowledgements}
This work is supported by the Deutsche Forschungsgemeinschaft via SFB-TR 24 (project A5), the North-German Supercomputing Alliance (HLRN) via grant shp0006, and the DAAD via the RISE programme. 
\end{acknowledgements}

\end{document}